\documentclass[a4paper]{jpconf}
\usepackage{graphicx}
\begin{document}
\title{The toolbox of modern multi-loop calculations: novel analytic and semi-analytic techniques}

\author{Alexey Pak
}
\address{
  Institut f{\" u}r Theoretische Teilchenphysik, 
  KIT,
  76128 Karlsruhe, Germany 
}
\ead{apak@particle.uni-karlsruhe.de}

\begin{abstract}
We describe three algorithms for computer-aided symbolic multi-loop calculations
that facilitated some recent novel results. First, we discuss an algorithm to
derive the canonical form of an arbitrary Feynman integral in order to facilitate their identification.
Second, we present a practical solution to the problem of multi-loop
analytical tensor reduction. Finally, we discuss the partial fractioning of
polynomials with external linear relations between the variables.
All algorithms have been tested and used in real calculations.
\end{abstract}

\section{Introduction}

A higher-order calculation is a multi-stage process, with details strongly dependent on
the physics problem. There is no generic one-size-fits-all approach, but usually one
first generates the diagrams, then performs Dirac algebra and/or projections on scalar
integrals, and then manipulates large expressions that depend on scalar products of
loop and external momenta until they can be reduced to known integrals and computed.
Each calculation has its own ``key'' stage which requires the most efforts. In many cases, 
those are the algebraic reduction of integrals (with integration-by-parts identities
~\cite{'tHooft:1972fi,Tkachov:1981wb,Chetyrkin:1981qh}) to a small
subset of ``master'' integrals, and the evaluation of the latter. Those problems have been the
focus of active research for many years and gave rise to a plethora of various methods
and algorithms~\cite{Smirnov:2006ry}. However, as the three-, four- and
higher-loop problems become the norm in the LHC era, even the traditionally ``simple''
steps become impossible to perform manually or with the ad-hoc methods.

In this contribution we present the three algorithms that automate the tasks usually
considered ``routine'' and not worth mentioning. First, we discuss the classification of
integrals as instances of different ``topologies'', or families of integrals that differ by the
exponents of the (fixed) denominator factors. When the number of such families reaches a
few hundred, automation becomes a necessity.

Second, we consider the tensor reduction, that may e.g. help re-arrange numerators of
factorizable topologies in order to perform integration independently. In certain cases,
involving complicated asymptotic expansions, this stage may dominate the total computation
time of a diagram. In addition, we provide a practical solution for the situations where 
general formulas are not known.
Finally, we present a (relatively straightforward) re-formulation of the partial fractioning
problem and solve it using Gr{\" o}bner bases.

The primary use of these methods is to generate the code for the computer algebra system
that processes the actual diagrams (e.g. FORM~\cite{Vermaseren:2000nd}), and thus the 
runtime efficiency is not the highest priority. Nevertheless, we find that our solutions are rather 
efficient and do not require enormous times for real-life problems.

\section{Identification of Feynman integrals}

In multi-loop calculations, one often has to solve a problem of identifying individual Feynman
integrals or deciding whether an integral belongs to a given family (``topology'') which helps 
reduce the number of integrals to compute. 
In general, in order to say whether two integrals are equal one has to compute both.
However, for some integrals it is possible to elaborate a simple transformation of loop 
momenta and establish the identity of integrands. In simple cases, such transformations
can be derived manually just by looking at graphs, but in multi-loop multi-scale problems
with non-trivial kinematic constraints this is a daunting task. 

\begin{figure}[b!]
  \hspace{0.1\textwidth}
  \includegraphics[width=0.3\textwidth]{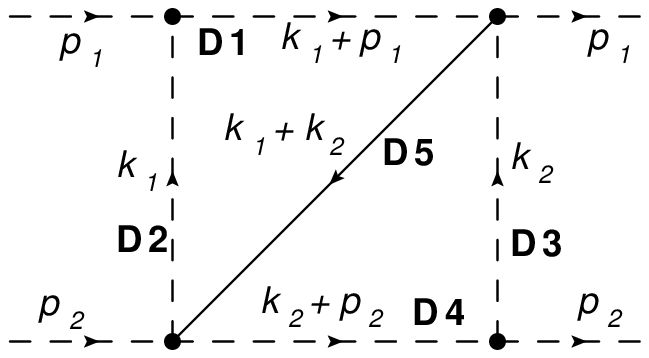}
  \hspace{0.2\textwidth}
  \includegraphics[width=0.3\textwidth]{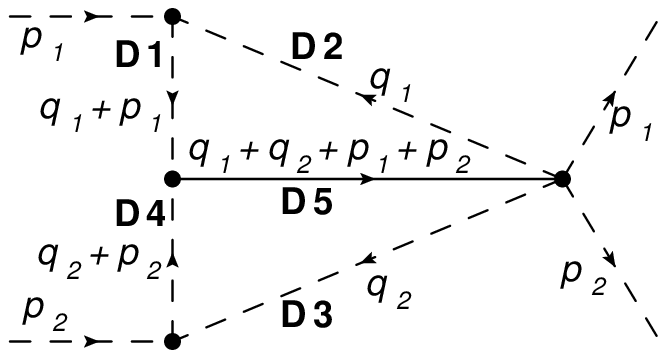}
  \hfill \\
  \caption{\label{f1} Graphs corresponding to integrals $I_1$ and $I_2$.}
\end{figure}

Let us consider two forward-scattering integrals (Fig.~\ref{f1}):
\begin{eqnarray}
  I_1 &=& \int \frac{d^D k_1~ d^D k_2}{D_1 D_2 D_3 D_4 D_5}, ~~
  \begin{array}{l}
    D_1 = (k_1 + p_1)^2, ~
    D_2 = k_1^2,~
    D_3 = k_2^2, \\
    D_4 = (k_2 + p_2)^2,~
    D_5 = (k_1 + k_2)^2 + m^2,
  \end{array} \\
  I_2 &=& \int \frac{d^D q_1~ d^D q_2}{E_1 E_2 E_3 E_4 E_5}, ~~ 
  \begin{array}{l}
    E_1 = (q_1 + p_1)^2, ~
    E_2 = q_1^2, ~
    E_3 = q_2^2, \\ 
    E_4 = (q_2 + p_2)^2, ~
    E_5 = (q_1 + q_2 + p_1 + p_2)^2 + m^2 
  \end{array}
\end{eqnarray}
with the common additional constraints $p_1^2 = p_2^2 = 0$, $(p_1 + p_2)^2 = -1$.

The integrands of $I_1$ and $I_2$ obviously cannot be related by only renaming
symbols. However, the value of e.g. $I_1$ is invariant with respect to a shift
of loop momenta, e.g. $k_1 \to k_1^\prime = k_1 + p_1$, or a permutation of 
denominator factors, e.g. $D_1 \leftrightarrow D_2$, etc. There may
exist a transformation $D_i(K) = E_{p(i)}(Q)$ with $K = M Q$ and  
$p(i)$  is some permutation of indices $1,2,...,5$, such that
\begin{eqnarray}
  K = \left(\begin{array}{c} k_1 \\ k_2 \\ p_1 \\ p_2 \end{array}\right),~~
  Q = \left(\begin{array}{c} q_1 \\ q_2 \\ p_1 \\ p_2 \end{array}\right),~~
  M = \left(\begin{array}{c|c} A & B \\ \hline 0 & I \end{array}\right),
\end{eqnarray}
$A$ and $B$ are the $2\times 2$ matrices,
$|A| = 1$\footnote{or in general $|A| = d \ne 0$, and integrals 
then are equal modulo corresponding Jacobian.}
, and $I$ is a $2\times 2$ identity matrix.

One practical way to compare two topologies is to compare graphs.
In Fig.\ref{f1} we present the graphs corresponding to $I_1$ and $I_2$,
and one can see that the graphs do not match.
However, some integrals may have no corresponding graphs (for example, 
integrals originating from some effective theories may have linear
denominators, e.g. $2 p_1 k_1 + i0$, which cannot be interpreted as graph lines),
and some integrals may have multiple corresponding graphs (as in the given
case). In addition, (sub)graph isomorphism is itself a computationally hard
problem\footnote{Subgraph isomorphism problem is NP-complete, while graph
isomorphism belongs to the NP complexity class.}.

Instead, in order to compare integrals, one could use alpha-representation~\cite{Bogolyubov:1980nc},
which is explicitly covariant, very closely related to the definition of
dimensionally regularized Feynman integrals and can be derived for any set of quadratic denominators.
The structure of any Feynman integral is encoded by the two homogeneous 
polynomials, $U$ and $F$ which do not depend on the exponents of denominator
factors and the dimensionality of space $D$.
In particular, we have:
\begin{eqnarray}
  && I_i = C \int d x_1~ ...~ d x_5~
    \delta\left(x_1 + ... + x_5 - 1\right)~
    U_i^a F_i^b, \\
  && U_1 = U_2 = x_5 (x_3 + x_4) + (x_1 + x_2)(x_3 + x_4 + x_5),  \nonumber \\
  && F_1 = m^2 x_5 U_1 - x_1 x_4 x_5,~~
  F_2 = m^2 x_5 U_1 - x_2 x_3 x_5. \nonumber
\end{eqnarray}
with constants $C$, $a$ and $b$ (irrelevant here) depending only on $D$.

The only freedom that is left when comparing $\{U_1, F_1\}$ with $\{U_2, F_2\}$
is a permutation of lines, or equivalently, variables $x_i$. One then has to find a
permutation $p(i)$ such that $U_1(\vec{x}_i) = U_2(\vec{x}_{p(i)})$ and 
$F_1(\vec{x}_i) = F_2(\vec{x}_{p(i)})$. Naturally, before looking at permutations, one
may compare more mundane properties of polynomials common to all permutations, 
such as the total number of terms or the number of terms proportional to $m^2$.

An obvious way to find $p(i)$ when comparing two integrals would be to try all $5! = 120$
permutations of $x_i$. Since usually one has more than two candidates to compare,
it is beneficial to introduce a ``canonical'' ordering,
maximizing some metric over all permutations. The canonical
ordering of lines should then be derived only once per integral, and the
identity of canonical alpha-representations would provide the definitive answer.

In practice, we found that it is sufficient to build a metric in the space of not pairs $\{U,F\}$ but
products $U F$. One example of a suitable metric is given by the following rules (we assume
that a unique ordering of coefficients is available, as it is in any computer algebra system):
\begin{enumerate}
\item{Turn the polynomial of $n$ variables with $m$ terms into a matrix with rows
  corresponding to monomials. The first column contains coefficients, and the subsequent
  columns contain the (non-negative integer) exponents of variables $x_1, ..., x_n$.}
\item{Make $n$ copies of the table. In the $i$-th copy, exchange the second column
  (corresponding to $x_1$) with the $i$-th column (originally corresponding to $x_i$).}
\item{In all copies, sort rows lexicographically by the first two columns (i.e. compare only the
  first two entries in each row).}
\item{Extract the second column from each copy (as vectors of length $m$), and determine
  the lexicographically largest vector, comparing all $m$ elements.}
\item{In the table copies with the maximized second column, continue recursively:
  produce $n-1$ copies-of-copies, in each select a different third column, sort
  by the three first entries, find the maximum third column, discard non-maximal entries, etc.}
\item{The permutations of columns in the copies maximizing all columns (there can be 
  a few due to symmetries) can be taken as the ``canonical permutations'' of $x_i$.}
\end{enumerate} 
\begin{figure}[t!]
  \hspace{0.1\textwidth}
  \includegraphics[width=0.3\textwidth]{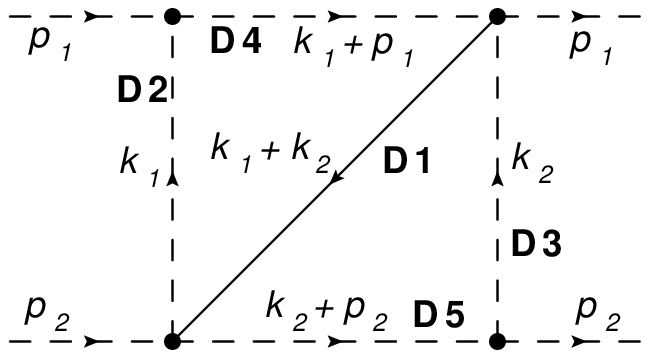}
  \hspace{0.2\textwidth}
  \includegraphics[width=0.3\textwidth]{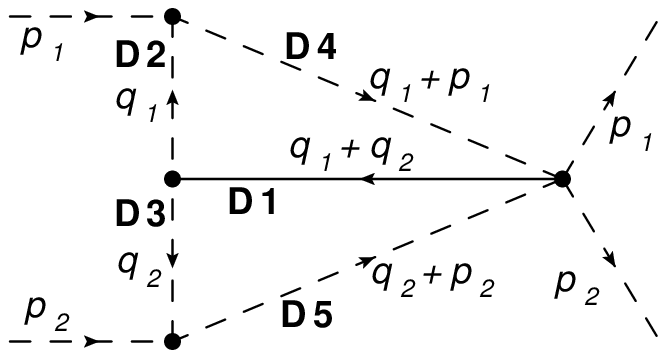}
  \hfill \\
  \caption{\label{f2}Graphs corresponding to integrals $I_1$ and $I_2$ after identical  loop re-parametrization.}
\end{figure}

While there is a theoretical possibility of combinatorial growth in this strategy, we have found
this approach quite fast and practical to at least five-loop integrals. In our example, after the
canonical re-ordering, one easily establishes that
$U_1 F_1 (x_5, x_2, x_3, x_1, x_4) = U_2 F_2 (x_5, x_1, x_4, x_2, x_3)$, 
i.e. $I_1$ and $I_2$ are indeed identical. After this identification it is easy to
find the relation between the loop momenta. The graphs, however, may differ as shown in Fig.\ref{f2}.

\section{Analytic tensor reduction}

A traditional approach to computing multi-loop integrals with tensor structures in numerators 
is to decompose the result into all tensors allowed by the symmetries and then determine 
the coefficients of the decomposition. Let us consider a simple one-loop example:
\begin{eqnarray}
  \label{e1}
  \int k^{\mu_1} k^{\mu_2} k^{\mu_3} k^{\mu_4} \cdot \frac{d^D k}{k^2 + m^2}
  = \langle k^{\mu_1} ... k^{\mu_4} \rangle
  = \sum_i C_i T_i^{\mu_1 ... \mu_4}. 
\end{eqnarray}
The tensors $T_i^{\mu_1 ... \mu_4}$ cannot depend on any external momenta, 
thus they may only be composed of metric tensors. There are only three unique products:
$T_1 = g^{\mu_1\mu_2} g^{\mu_3\mu_4}$,
$T_2 = g^{\mu_1\mu_3} g^{\mu_2\mu_4}$, and
$T_3 = g^{\mu_1\mu_4} g^{\mu_2\mu_3}$.

The coefficients $C_i$ can be found by multiplication of the original equation Eq.~\ref{e1}
with $T_1$, $T_2$, and $T_3$ and solving the system of equations:
\begin{eqnarray}
  && {T_1}_{\mu_1 ... \mu_4}  \langle k^{\mu_1} ... k^{\mu_4} \rangle
  = \langle (k^2)^2 \rangle = C_1 D^2 + C_2 D + C_3 D, \\ \nonumber
  && {T_2}_{\mu_1 ... \mu_4}  \langle k^{\mu_1} ... k^{\mu_4} \rangle
  = \langle (k^2)^2 \rangle = C_1 D + C_2 D^2 + C_3 D, \\ \nonumber
  && {T_3}_{\mu_1 ... \mu_4}  \langle k^{\mu_1} ... k^{\mu_4} \rangle
  = \langle (k^2)^2 \rangle = C_1 D + C_2 D + C_3 D^2
\end{eqnarray}

The equality of all products on the left hand side implies that
$C_1 = C_2 = C_3 = C$, thus reducing the number of independent
equations to one, and we find $C = \left[D (D+2)\right]^{-1} \langle (k^2)^2\rangle$.

In this case, it is not difficult to generalize the formula to a general number of indices:
the result will be represented in terms of symmeterized product of metric
tensors, and the coefficients will contain a number of gamma-functions. 
In a higher number of loops, however, there exist only a few general formulas.
The two-loop generalization of Eq.\ref{e1}, first given in ~\cite{Davydychev:1995nq}
involves Gegenbauer polynomials, traceless tensors, and is much more
computationally expensive due to multiple nested summations.
Tensor reduction may also be derived in terms of dimension shifts and differential
operators~\cite{Tarasov:1998nx}, but that is not very convenient in real calculations.

Instead, we suggest to directly use the method as outlined above for the fixed number
of indices. For illustration, let us consider a two-loop propagator-type integral with loop
momenta $k_1$ and $k_2$, and the external momentum $p$,
where the formulas of Davydychev-Tausk still apply (although with some effort),
and limit ourselves with six open indices in the numerator.

In the left-hand side we can have seven distinct distributions of indices over the
loop momenta (e.g., 
$\langle  k_1^{\mu_1}  k_1^{\mu_2}  k_1^{\mu_3}  k_1^{\mu_4}  k_2^{\mu_5}  k_2^{\mu_6}\rangle$).
On the right-hand side we may build 76 possible tensors out of metric tensors and the
components of the external momentum $p$ (e.g., $g^{\mu_1\mu_2} p^{\mu_3} g^{\mu_4\mu_5} p^{\mu_6}$).

For each of the seven LHS combinations, the coefficients in front of the RHS tensors have to be determined separately.
Naively, in every case one would need to solve 76 equations with 76 variables.
However, the example above gives us a hint: one may reduce the number of independent
variables by exploiting the symmetries. If the products of an LHS tensor with the two RHS tensors
coincide, so will the corresponding coefficients. In this case, one has at most 10 independent
coefficients to determine (each time from 76 equations).

This procedure does not depend on the actual denominators of the integral: we only derive a
decomposition of the numerators. Thus, it is possible to pre-compute the tables with
the reductions for a sufficiently large number of indices and re-use them in different computations.
One only needs a practical way to solve the systems of linear equations,
where coefficients depend on dimensionality $D$ and possibly kinematic invariants.
In our setup, we reuse the components from the Laporta algorithm that perform Gauss
reduction.

We have checked that this approach is indeed very practical: for example, our implementation
builds a table for all 4-loop propagator-type integrals with up to 6 indices in the numerator in
less than one hour with a regular PC.

\section{Partial fractioning}

\begin{figure}
  \includegraphics[width=0.3\textwidth]{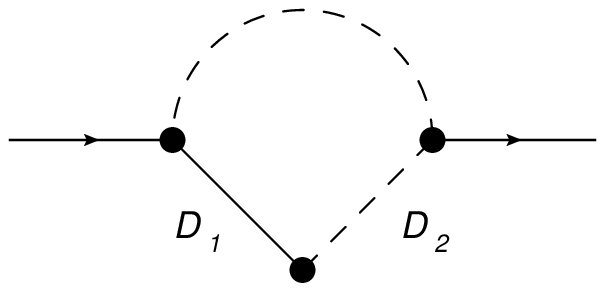}
  \put(-30,0){(a)}
  \hfill
  \includegraphics[width=0.3\textwidth]{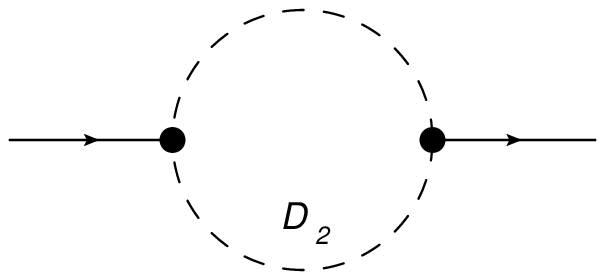}
  \put(-30,0){(b)}
  \hfill
  \includegraphics[width=0.3\textwidth]{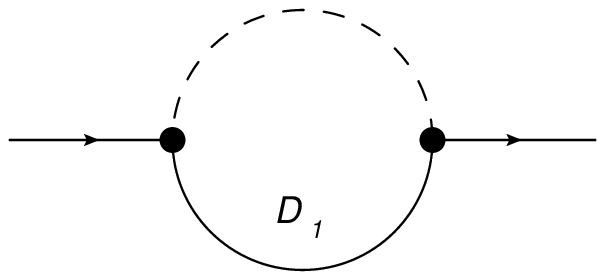}
  \put(-30,0){(c)}
   \\
  \caption{\label{f3} Original (linearly dependent, (a)) and fraction-decomposed ((b) and (c)) topologies.}
\end{figure}

Normally after the reduction to scalar integrals one has a long polynomial with terms
that differ in exponents of denominator factors, e.g.
\begin{eqnarray}
  \label{e1}
  &&{\rm diagram}
    = D_1^2 D_2 + 33 D_1^{-1} D_2^{-2}
    - 45/16 D_1 D_2^{-1} + (\epsilon + 2) D_2 + ... , \\ \nonumber
  &&
  D_1 = k^2 + m^2, ~~
  D_2 = k^2, ~~\mbox{i.e. linear dependence:}~~ D_1 - D_2 - m^2 = 0.
\end{eqnarray}
(the corresponding graph is in Fig.~\ref{f3} (a)). In what follows we will discuss
general transformations of monomials $D_1^a D_2^b$, where $a$ and $b$ are positive 
or negative integer numbers.

The condition that $D_1$ and $D_2$ are dependent allows us to simplify such monomials.
For example, we may use the ``classical'' fraction decomposition relation
\begin{eqnarray}
  \label{e2}
  \frac{1}{D_1 D_2} \to \frac{1}{m^2 D_2} - \frac{1}{m^2 D_1}
\end{eqnarray}
repeatedly to reduce any monomial with negative exponents $a$ and $b$ to a linear 
combination of terms where at least one of $a$ or $b$ is non-negative. 

The form where one of $D_1$ and $D_2$ is absent is preferable for the further computations.
In this case, the goal of the fraction decomposition is to identically re-arrange Eq.~\ref{e1} so that
each term in the final expression could be identified with one of the ``simpler'' topologies 
in Fig.~\ref{f3}(b) and (c).

However, the rule Eq.~\ref{e2} alone does not achieve this goal. In particular, it does not apply 
when $b > 0$, $a < 0$ and one has to apply the direct decomposition rule $D_2 \to D_1 - m^2$.
In the symmetric case $b < 0$, $a > 0$ it could be convenient to use $D_1 \to D_2 + m^2$, but
applying both those rules to monomials with $a > 0$ and $b > 0$ will not terminate. After the 
inspection of all relevant cases one can find that that only a system of three rules
\begin{eqnarray}
  \left(D_1 D_2\right)^{-1} \to \left(m^2\right)^{-1} \left(D_2^{-1} - D_1^{-1}\right),~~
  D_2 \to D_1 - m^2,~~
  D_1/D_2 \to m^2/D_2 + 1
\end{eqnarray}
provides the complete decomposition.

\begin{figure}
  \centering
  \includegraphics[width=0.4\textwidth]{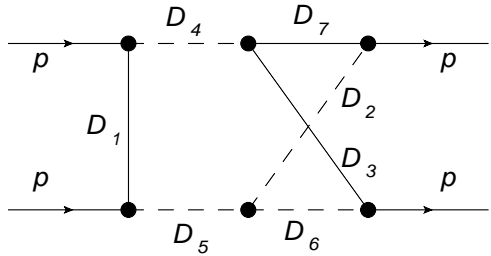}
  \caption{\label{f4} Linearly dependent topology with two relations between denominator factors.}
\end{figure}

The example above is relatively straightforward and can be easily done manually.
However, in practice one may have a much more complicated case for linearly
dependent factors. In Fig.~\ref{f4} is a two-loop example of a topology in special
kinematics, where $p^2 = - m^2$ and each massive line has mass $m$. Here one
has two relations between the seven factors:
\begin{eqnarray}
  \label{e3}
  && 2 D_1 - D_4 - D_5 - 4 m^2 = 0, \\ \nonumber
  && 2 D_2 - 2 D_3 + D_4 - D_5 + 2 D_6 - 2 D_7 + 4 m^2 = 0.
\end{eqnarray}
Let us formulate the problem in a slightly different way. First, let us get rid of the
negative powers of $D_i$ by introducing variables $Y_i = D_i^{-1}$. Second,
for all monomials $D_1^{a_1} ... D_2^{a_7} Y_1^{a_8} ... Y_7^{a_{14}}$,
$a_j \ge 0$, let us introduce an ordering that incorporates our understanding of
what is ``simpler''. In particular, it is beneficial to use linear weighting of exponent 
vectors $(a_1,...,a_{14})$.
Given two such vectors $\vec{a} $ and $\vec{b}$, we decide which of them is 
the ``largest'' by lexicographically comparing the products $M\vec{a}$ and $M\vec{b}$,
where $M$ is some square $14 \times 14$ matrix. One reasonable choice is
$M_{ij} = \left\{\begin{array}{l}1,~~ {\rm if}~~ i \ge j \\ 0, ~~{\rm otherwise} \end{array}\right.$.
The sum of all exponents is here the primary criterion, which agrees with the intuitive 
definition that the monomials with fewer non-zero exponent are ``simpler''.

Given this ordering, the problem is to re-arrange polynomial conditions Eq.~\ref{e3} and the
additional relations $D_i Y_i - 1 = 0$ into a set of equivalent relations that would unambiguously
reduce any given monomial to the ``simplest'' form (and would not lead to loops during
the repeated application).

As formulated above, exactly this problem is solved by the so-called Buchberger algorithm, 
that generates a set of polynomials known as the Gr{\" o}bner basis. We then need to interpret
each element of this basis (a polynomial $p = 0$) as a rule to substitute its most ``complex''
monomial (according to ordering $M$) with the remaining terms. Quite conveniently,
Mathematica has a function \verb+GroebnerBasis[...]+ that is rather efficient and has an
option \verb+MonomialOrder+ to select the proper weight matrix $M$.
In this particular case, this function produces a Gr{\" o}bner basis
consisting of 14 polynomials that would be very difficult to derive manually.
This output can be directly translated to FORM code and used to decompose expressions of
any complexity.

\subsection{Acknowledgements}

The author is indebted to A. Smirnov, K. Chetyrkin and M. Steinhauser for useful discussions.

\section{References}
\bibliographystyle{iopart-num}
\bibliography{apak}

\end{document}